\documentclass[preprint]{elsarticle}
\usepackage{natbib}
\usepackage{lineno}
\usepackage{amsmath}
\usepackage{indentfirst}
\usepackage{amsfonts}
\usepackage{amssymb}
\usepackage{changepage}
\usepackage{lscape}
\usepackage[]{algorithm2e}
\usepackage{afterpage}
\usepackage[T1]{fontenc}
\usepackage{booktabs}
\usepackage{url}
\usepackage{comment}

\usepackage{subfigure}
\usepackage{color}
\usepackage{epsfig}
\usepackage{epstopdf}
\usepackage{amsmath}
\usepackage{graphicx,psfrag,epsf}
\usepackage{afterpage}
\usepackage{color}
\usepackage{soul}
\usepackage{multirow}
\usepackage{rotating}
\usepackage{scrextend}

\newcommand{\bX}{\mathbf{X}}

\newcommand{\by}{\mathbf{y}}

\newcommand{\Expect}{{\rm I\kern-.3em E}}

\newdefinition{defin}{Definition}

\begin{document}
\begin{frontmatter}
\title{\textbf{A Set of Rules for Model Validation}}

\author{Jos{\'e} Camacho\fnref{fn1}}
\ead{josecamacho@ugr.es} 
\address{University of Granada}
\fntext[fn1]{Signal Theory, Networking and Communications Department, University of Granada, C/ Periodista Daniel Saucedo Aranda s/n 18071, Granada, Spain}

\date{}

\begin{abstract} 
The validation of a data-driven model is the process of assessing the model's ability to generalize to new, unseen data in the population of interest. This paper proposes a set of general rules for model validation. These rules are designed to help practitioners create reliable validation plans and report their results transparently. While no validation scheme is flawless, these rules can help practitioners ensure their strategy is sufficient for practical use, openly discuss any limitations of their validation strategy, and report clear, comparable performance metrics.  
\end{abstract} 
\end{frontmatter}   

\section{Introduction}  \label{Introduction}

Model validation is a fundamental task in all modern data-driven systems, whether they fall under the broad categories of Statistics, Machine Learning, Artificial Intelligence, or more specialized fields like chemometrics. Validation has become a major focus for regulatory and standardization bodies, with key reports and standards highlighting the growing concern for ensuring the trustworthiness and reliability of data-driven models:

\begin{itemize}
    \item NIST Artificial Intelligence Risk Management Framework (AI RMF 1.0, 2023): Published by the U.S. Department of Commerce, this framework provides management techniques to address the risks and ensure the trustworthiness of AI systems, with validation as a core component.
    \item The EU AI Act of 2024, landmark piece of EU legislation that categorizes AI systems by risk level, where validation is not defined as a best practice but a legal requirement within the conformity assessment.
    \item The ISO/IEC TS 4213:2022, by the  International Organization for Standardization (ISO), describes approaches and methods to ensure the relevance, legitimacy and extensibility of machine learning classification performance assertions.
    \item  The American Society for Testing and Materials (ASTM) E2617 provides a formal framework for the validation of empirically derived multivariate calibrations/models.
    
\end{itemize}

Besides these official bodies, there has been a renewed effort in scientific journals to discuss and promote sound practices for model evaluation and validation in data-driven research~\cite{raschka2018model,xu2018splitting}, which resonates in the field of chemometrics~\cite{ezenarro2025xyonion,lopez2023importance,fearn2025multivariate,ezenarro2025chemometric}.

All previous literature is either regulatory or methodological, focusing on a list of specific data processing methods and good practices, like the use of external validation and/or cross-validation. The approach of this perspective paper is different: to suggest a reduced number of conceptual, general rules for proper validation. These rules provide the underlying intuition for the methods discussed elsewhere. Consequently, they can be leveraged to understand when and why these methods are relevant, or to derive new validation techniques when they are not. In particular, the rules are meant to be useful for:

\begin{itemize}
    \item Designing sound validation schemes in complex problems/data pipelines\footnote{In this paper, "data pipeline" refers to the set of processing steps performed on the data for its analysis and modeling (e.g., data cleaning, preprocessing, analysis, modeling)}.
    \item Reporting transparent and comparable performance results.
    \item Specifying limitations of reported validation results, in terms of different forms of risks.
\end{itemize}

A core principle of these rules is the recognition that model validation is rarely perfect, so risks need to be reported along with performance evaluation results.

\section{Rules of Validation}

We need to start with the definition of the term validation. In the field of classical statistics, validation is often interpreted as the evaluation of the stability and statistical significance of the
model parameters \cite{pena2014fundamentos}. Thus, the model is valid if the parameters are statistically significant---and therefore unlikely to be the result of a purely random process---and stable, meaning they would not change dramatically in response to expected variations in the input. This paper, however, focuses on the more popular definition of validation in the field of machine learning, interpreted as the
evaluation of the model's capability for performance generalization. A valid model is the one that maintains a high modeling performance when applied to new independent data from the population of interest, that is, the potentially unlimited set of data objects for which the model makes some form of inference---prediction, classification, hypothesis test, anomaly detection, etc. This second definition also reflects the relevance of stability when applied to new data, but substitutes statistical significance with performance generalization. Note that these three interconnected concepts: stability, statistical significance and performance generalization, have a principal role in the proposed validation rules.

\subsection{Rule 1: Use \textbf{independent} data for model \textbf{building} and for the evaluation of the \textbf{generalization performance} }

Data-driven models have two types of parameters: the regular ones, which values are automatically derived from data using a fitting algorithm, and the meta-parameters, which values are selected by the analyst in a pseudo-automatic fashion. For instance, in Principal Component Analysis (PCA) we may use the NIPALS algorithm to fit loadings and scores (the regular parameters), and cross-validation to select the number of components (the single PCA meta-parameter) \cite{wold1978cross,bro2008cross}. Thus, \textbf{model building} comprises two interrelated tasks: 

\begin{itemize}
    \item \textbf{Training or data fitting}. Both terms are interchangeably employed by the Machine Learning community, while in chemometrics the term 'calibration' is po\-pular due to the wide interest in extracting quantitative information from complex analytical instruments, which have to be calibrated. Training, data fitting or calibration refer to the estimation of regular parameters using a suitable fitting algorithm. 

    \item \textbf{Model selection}. From the set of model variants with different meta-parameters, select the model with the best performance.
\end{itemize}

The whole process of model building can be done with two data splits (often called ‘training’ and ‘validation’ sets) or through some form of cross-validation. Once the model is built, the \textbf{generalization performance} is then evaluated with an additional ‘test’ set or through double/nested cross-validation~\cite{filzmoser2009repeated}. The separation between model building and test data is a cornerstone of the engineering of data-driven models, since there is often a significant gap between performances in validation and test sets~\cite{xu2018splitting}, so that performance is often higher in the former. When such a gap exists, we say that the model is overfit to the model building data, capturing patterns that are only found in these data but do not generalize to the test set and to the population of interest.

The first rule states that the 'test' set needs to be drawn \textbf{independently} of the dataset(s) used during model building. The problem lies in the definition of independence, which is more intricate than one might initially expect. This issue is treated in detail in the next rule. The consequence of a lack of independency is that the \textbf{perceived generalization performance} of the model gets too optimistic, appearing that the model performs better than what it actually does. Notice here an important distinction: the perceived generalization performance is the one inferred by applying the model to the test set, while the true generalization performance is the one for the model applied over the population of interest.

The dependency between model building and test data constitutes one of the major risks in validation and is one type of the so-called \textbf{data leakage}, which happens when a model is trained using information that would not be avai\-lable at the time of real-world practice. This dependency inflates the perceived generalization performance, because it incorporates patterns that are true for both the model building data and the test data, but are not generally true for the population of interest. Those patterns constitute the leakage information.

A final remark is that, strictly speaking, the ‘training’ and ‘validation’ sets do not need to be independent. Actually, some fast model selection procedures may indeed violate the independence between ‘training’ and ‘validation’ data~\cite{saccenti2015use}.

\subsection{Rule 2: the \textbf{test set}, the \textbf{population of interest} and the \textbf{real-life application} of the model need to be consistent}

As Esbensen and Geladi claim~\cite{esbensen2010principles,LopezCorrillero2023} "All prediction models must be validated with respect to realistic future circumstances". This statement is particularly relevant in terms of the completeness of the test data and its level of independence with the model building data, among other considerations that affect data quality~\cite{wasielewska2022evaluation}.

%Concepto complicado, me lo dejo para el futuro: Both are interrelated concepts: the level of independence reflects the number of real factors that are considered between model building data and test data (e.g., laboratorio, tecnico, etc.) and the completeness correspond to the amount of levels in those factors.  

\textbf{If the test set is poorly representative of the population of interest, the perceived generalization performance will be of little practical use}. This is where concepts like completeness and bias are relevant. For instance, when we build models for disease diagnosis in humans using measurements over blood samples, we would ideally like those models to work for the entire population of blood samples coming from humans, regardless of sex, race, geographical location, biological predisposition, exposures, the laboratory where samples were measured, the measurement protocol, etc. Unfortunately, it is difficult, expensive or even unrealistic to gather a sufficiently representative test set taking all previous considerations and more into account. For this reason, compromises are most often made to derive simpler and less expensive models for the application at hand. For that purpose it is key to determine a practical definition of the population of interest and subsequently select a representative test set.

Selecting the level of independence of the test set is also complex because there are \textbf{sources of interdependence that can easily go unnoticed or are complex to avoid in real practice}. Dependency is inherently connected to incompleteness, since one form of dependency is the direct consequence of the latter. Take for instance the popular criticism of genomics models being biased due to the over-representation of European-ancestry populations~\cite{fatumo2022diversity}. When we build and test models mainly on these populations, we cannot expect that they generalize well to other human populations. First, because the test is incomplete, and so we do not know how the model performs for other types of samples. Second, because both model building and test data are incomplete in the same way, which creates a dependency that inflates the perceived generalization performance. 

\textbf{The requirements of the validation depend on the purpose of the model}, so that the completeness of the test set and the level of independence between model building and test data should be designed to mimic the real-life application. This is a powerful approach to design \textbf{sound and practical validation schemes}. Basically, we consider the practicalities of the application to decide the characteristics of the test set and the validation process. For instance, to improve local health practice in a hospital, it may be enough to validate a model in two alternative cohorts from that same hospital. 

An important element of a validation strategy that mimics the real-life application is that \textbf{any relevant data processing operation} during model building (either at model fitting or model selection) should \textbf{not consider the 'test' data} in any way possible. This is simply because in real-life practice, the data that will be inputted into the model is not available in advance, and test data should have the same consideration of no availability. Processing operations at model building which should be blind to the test set often include preprocessing options like mean-centering or auto-scaling (where means and scales can be considered model parameters), variable selection (selected variables can be considered meta-parameters), and similar operations~\cite{lopez2023importance}. This also means that nested cross-validation routines should be designed to perform these operations within the inner loop, without using the corresponding 'test' split of data. An unfortunately extended bad practice~\cite{lopez2023importance,ezenarro2025chemometric} is applying variable selection using the entire dataset and subsequently split the data in model building and test. An equivalent error is to perform variable selection and subsequently perform inferential procedures (like t-tests, ANOVA tests, univariate estimates of precision or recall, ROC curves, etc.) on the selected variables, since these procedures do not consider that the variables have been pre-selected from a larger group as the most promising ones.    

We should not confuse the previous situation with the one in which the data is \textbf{transformed with internal statistics}, for instance when working on normalized observations (e.g., for compositional data or derivatives in spectra) or on normalized windows of time series data (e.g., working on mean-centered intervals or incremental data). Such operations require to estimate statistics from each piece of data to preprocess that same piece of data. Because in the real-life application, each piece of data will be preprocessed with its own estimates, we can do the same for the 'test' set. 

Each real-life application involves convoluted practicalities relevant to validation. That is why it is valuable to reflect on the information that you have available in real practice and, more importantly, the information that you do not have available to design the validation approach. In the following, some examples to illustrate the rule criterion are discussed:

\begin{itemize}
    \item If the complete data is preprocessed variable-wise (e.g., data mean-centering), the test set for validation should be transformed with the preprocessing parameters estimated at model building.
    \item If pieces of data are preprocessed internally (e.g., mean-centering of intervals of time), the test set for validation is transformed with the preprocessing parameters estimated from itself.
    \item If there is potential lab/technical/batch effect and the model is meant to be valid in general practice, the test set should be generated at/by a different lab/technician/batch.
    \item If data are time series, the test set should be built from data collected in a different time interval than the model building data, separated enough in time to avoid autocorrelation between both datasets~\cite{bergmeir2018note}. Data shuffling is not consistent with real-life practice in time series. 
    \item If data is gathered from a time-varying process (i.e., a process that changes in time and so requires model updates), the test data should be separated in time from the model building data by the period that the model is expected to be operative without updating. 

\end{itemize}

Quite often, the data used in a study is provided as a single block and we resort to \textbf{data splitting}~\cite{lopez2023importance} for the generation of a model building dataset (potentially divided in training and validation) and a test dataset. Single and nested cross-validation is a form of data splitting. Please, note that in this situation, the test data is only independent to the model building data in terms of all the data processing operations that we make AFTER data splitting, but it is not independent in terms of all the previous operations, included the ones involved in the data generation process. This is a limitation that should be recognized.

Note that there are additional considerations to take into account, which can be derived from the second rule. Raschka~\cite{raschka2018model} states that data splits completely at random violate the statistical independence of the test set. This dependency comes from the fact that when classes or types of observations are split in a random, unbalanced manner, imbalance is reflected oppositely in building data and test data. This opposite relationship is not consistent with real-life practice, and can  produce a dramatic bias in the performance generalization estimates for small datasets and/or unbalanced classes. Raschka also claims that “stratified divides”, where all classes are randomly divided in model building and test,
should be preferred. Yet, stratification in the classes is not enough when observations are not independent in the real-life application (like in time series or repeated measures/replicates)~\cite{lopez2023importance}, and this dependency needs to be canceled among splits (e.g., avoiding auto-correlation among splits or including all replicates/repeated measures in the same split). On the other hand, Xu and Goodacre~\cite{xu2018splitting} claim that “systematic sampling” (e.g., Kennard-Stone) provides “poor estimation of the model performance”. Clearly, no systematic sampling is inspired in the real-life practice, and so its use needs to be carefully justified and ideally combined with other methods.

We may conclude that the more the validation strategy departs from the real-life application, the greater the risk of overestimating (or, more rarely, underestimating) the performance of the model. For proper transparency, the completeness and the level of independence at validation, e.g., data was analyzed in the same lab, using the same analytical instrument, in the same batch, etc., should be reported along with the validation results.

\subsection{Rule 3: The \textbf{criterion} for performance evaluation should be \textbf{objective} and \textbf{consistent} with the real-life application}

There is a large number of criteria for performance evaluation~\cite{ballabio2018multivariate}, to give some few examples: Predictive Error Sum-of-Squares (PRESS), the Goodness of Prediction ($Q^2$), Mean Absolute Error (MAE), Mean Square Error (MSE), Precision, Recall, F1 Score, Area Under the Receiver Operating Characteristic (AUROC), Number of Misclassification (NMC), Mathews Coefficient Correlation (MCC), Cohen’s Kappa, etc. Quite often, when comparing multiple model variants, these criteria provide conflicting results. In real-life applications, however, there are considerations to make, often of practical nature, about when a solution is preferred over other solutions. For instance, there may be different perceived degrees of severity when making false positives and false negatives. In the diagnosis of a severe disease, a false positive may be an unpleasant experience for the patient, but a false negative may lead to very negative consequences for their health. In a critical infrastructure, a false positive of malfunctioning may lead to wasting the time of a technician, but a false negative may lead to the disruption of the critical service. Even two different types of false positives (or negatives) may not be perceived the same in a real application. Understanding the severity of inference errors in the real-life application of a model is extremely important for proper validation. The application experts should provide this information. 

On the other hand, some criteria may be less suitable for specific data characteristics. For instance, the F1 Score is preferred to Precision and Recall for unbalanced data. ROC curves and AUROC are also affected by imbalanced data. 

\begin{figure}[htbp]
    \centering
 	\subfigure[30\% minority class]{\includegraphics[width=0.45\textwidth]{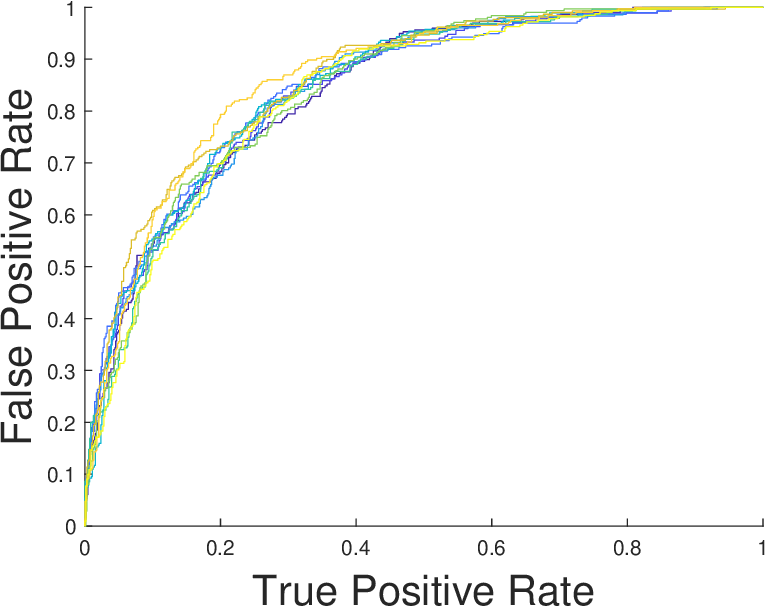}}
 	\subfigure[1\% minority class]{\includegraphics[width=0.45\textwidth]{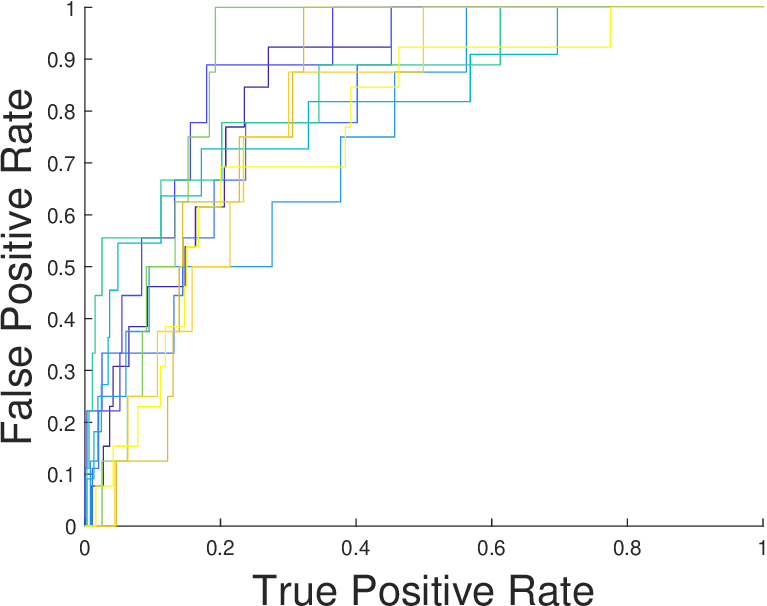}}
    \caption{Receiver Operating Characteristic (ROC) curves for unbalanced data. A ROC  curve is a graphical plot that illustrates the diagnostic ability of a binary classifier system as its discrimination threshold is varied, created by plotting the True Positive Rate (Sensitivity) against the False Positive Rate (1 $-$ Specificity) at various threshold settings.}
    \label{fig:ex1}
\end{figure}

Take the example of Figure \ref{fig:ex1}, where a random $\by~(1000 \times 1)$  is estimated by a hypothetical classifier $\hat\by~(1000 \times 1)$ with some level of noise. We repeat the generation of $\by$ and $\hat\by$ ten times, emulating what would happen if we assess the performance of the classifier over ten different test datasets with the same imbalance level. When $\by$ is composed of a majority class with $70\%$ of the observations and a minority class with $30\%$, the resulting ROC curve is very stable (Figure \ref{fig:ex1}(a), with an AUROC\footnote{The Area Under the ROC curve or AUROC ranges from 0.5 for a random classifier to an optimal value of 1 for a perfect classifier.} between 0.84 and 0.86). When we assess the same classifier with a minority class of $1\%$, which is realistic or even large for many anomaly detection problems or in the diagnosis of many diseases, the random variability associated to the ROC curve grows severely (Figure \ref{fig:ex1}(b), with an AUROC between 0.77 and 0.89). Clearly, a ROC (or AUROC) is an unstable performance measure for this type of problems.  

\begin{figure}[htbp]
    \centering
 	\subfigure[Minority 30\%]{\includegraphics[width=0.3\textwidth]{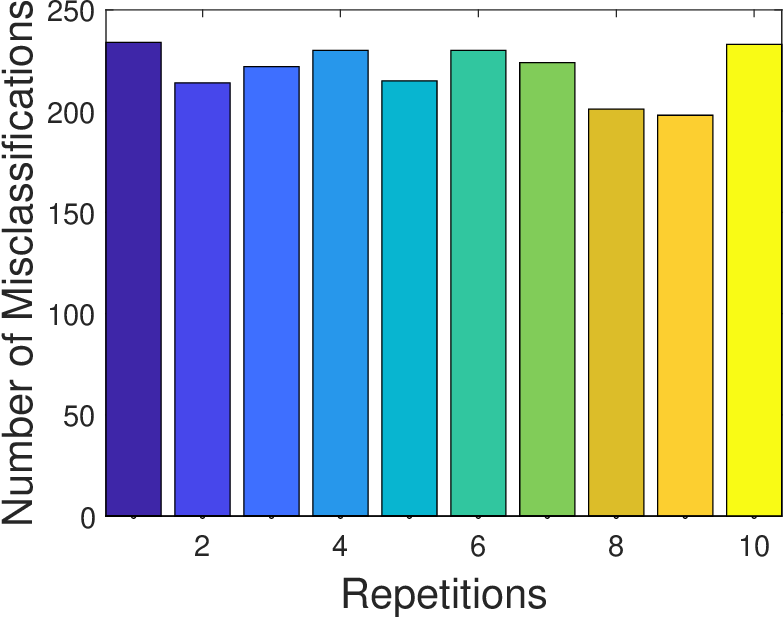}}
 	\subfigure[Minority  1\%]{\includegraphics[width=0.3\textwidth]{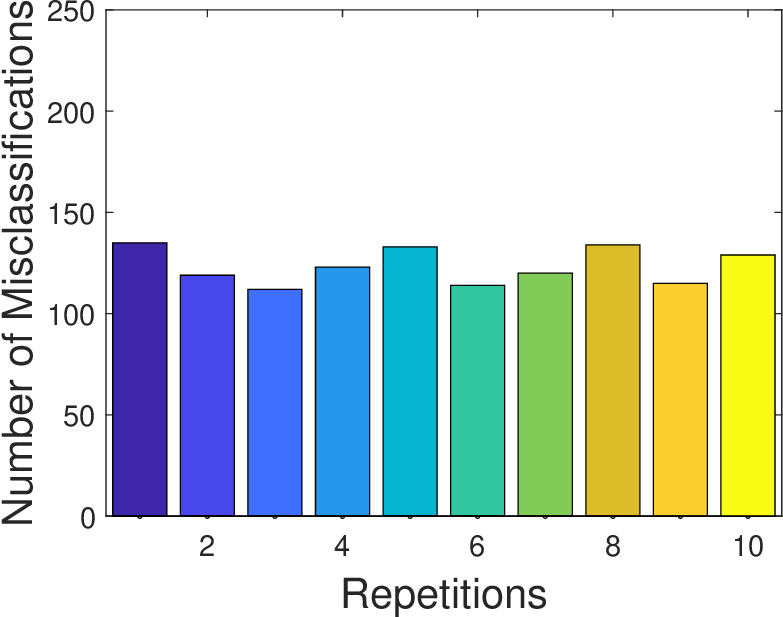}}
 	\subfigure[Minority 1\%, always negative]{\includegraphics[width=0.3\textwidth]{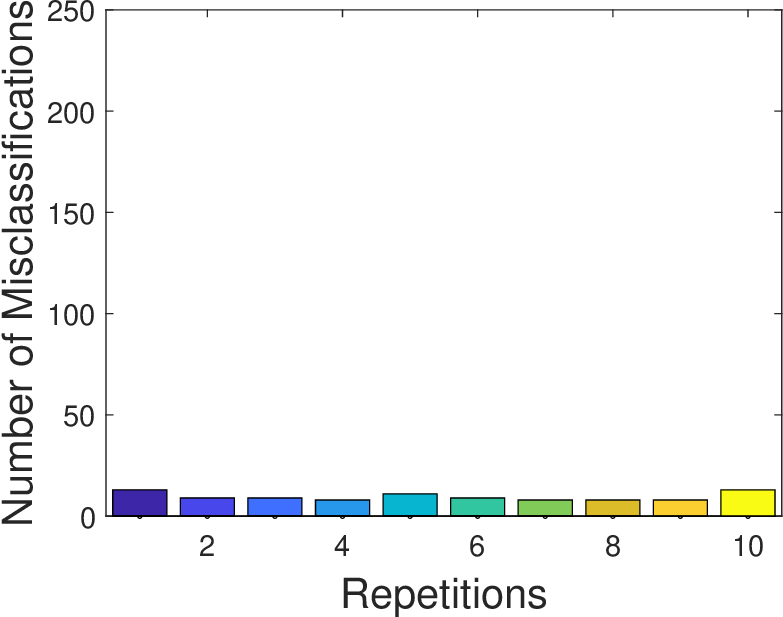}}
    \caption{Performance results based on the Number of Misclassifications (NMC) for unbalanced data. }
    \label{fig:ex1b}
\end{figure}

Another example is provided in Figure \ref{fig:ex1b} for the NMC and for the previous unbalanced cases. The NMC of the classifiers in Figures \ref{fig:ex1}(a) and \ref{fig:ex1}(b) are presented in Figures \ref{fig:ex1b}(a) and \ref{fig:ex1b}(b), respectively. The NMC is more stable than the ROC curve in the case with a minority class of $1\%$ (between 112 and 135), quite comparable in stability to the case with a minority class of $30\%$ (between 198 and 234). Yet, it is interesting to reflect on this result. The ROC curves in Figure \ref{fig:ex1}(b), even if unstable, show a good performing classifier, well outperforming the random classifier represented by an AUROC of 0.5. However, in terms of NMC and considering that we have only 1\% of positives, the classifier of Figure \ref{fig:ex1b}(b) performs worse than the naive approach that systematically outputs the negative class, shown in Figure \ref{fig:ex1b}(c). Given that the latter is clearly useless, we can conclude that the NMC, like it happens to the ROC curve, provides a flawed perception of the model performance for severely unbalanced data.  

\begin{figure}[htbp]
    \centering
 	\subfigure[0.99 Threshold]{\includegraphics[width=0.3\textwidth]{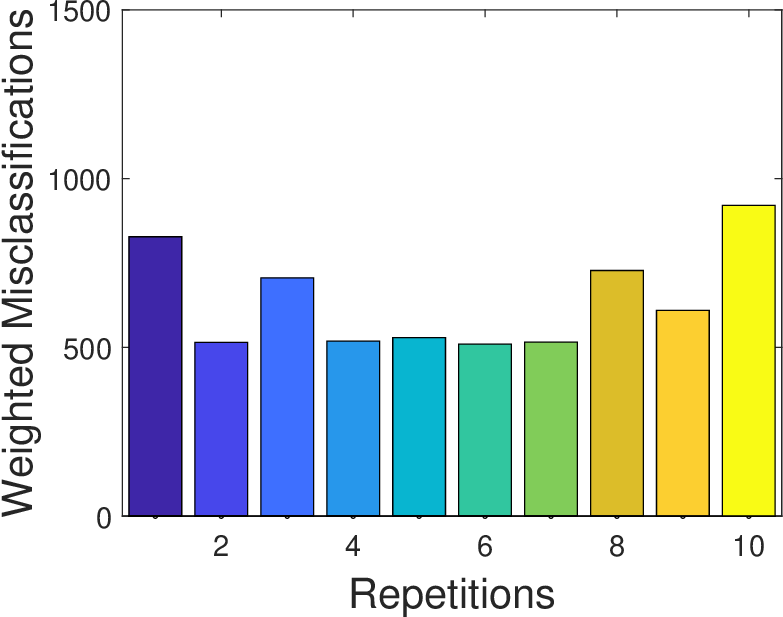}}
 	\subfigure[WMC Always negative 1\%]{\includegraphics[width=0.3\textwidth]{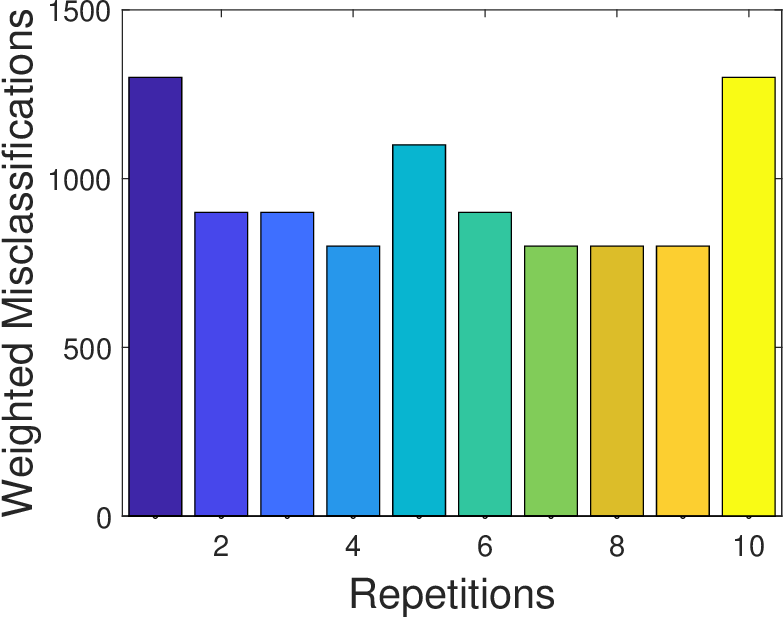}}
 	\subfigure[WMC 0.7 Threshold]{\includegraphics[width=0.3\textwidth]{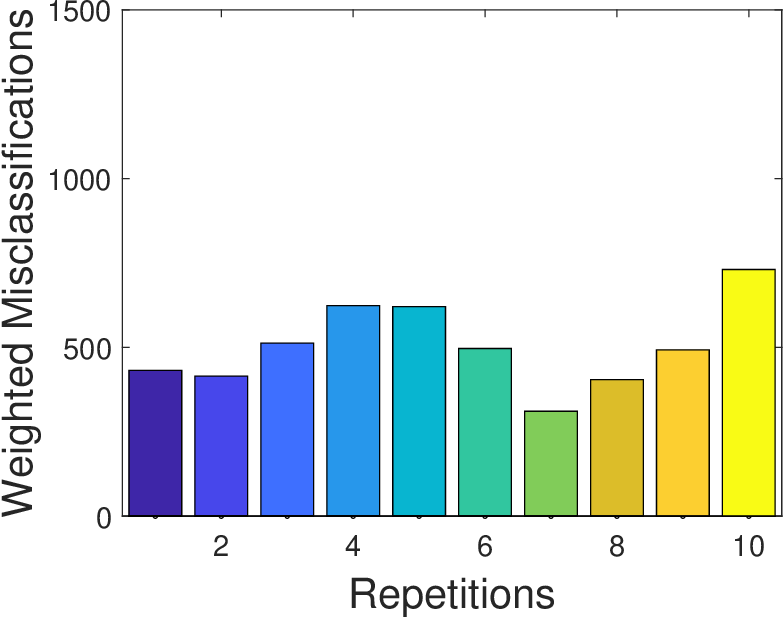}}
    \caption{Performance results based on the weighted number of misclassifications for unbalanced data with a minority class of 1\%.}
    \label{fig:ex1c}
\end{figure}

A different angle is considered in Figure \ref{fig:ex1c}. Rather than the NMC, we consider a weighted performance function where false positives are pondered with a unit and false negatives with 100 units, incorporating the idea that a false negative is something we would rather avoid at the expense of making more false positives. Back in the case with a minority class of $1\%$, this approach shows a better performance of the original classifier (in Figure \ref{fig:ex1c}(a)) than the naive approach that always outputs the negative class (in Figure \ref{fig:ex1c}(b)). This perceived performance is more consistent for a real-case problem where false negative are more relevant. Furthermore, this performance function shows that reducing the threshold for the detection (which generates less false negatives at the expense of generating more false positives) can further improve performance (Figure \ref{fig:ex1c}(c)). Clearly, the nature of the performance function can change both our perception of performance and the consequent choices we make to configure the optimal model. For this reason, choosing a performance criterion that is consistent with the real-life application is key to make adequate choices and have realistic assessment of the model quality.

The performance evaluation criterion (or criteria) should be reported along with a discussion of its suitability to consider the practicalities of the real application.

\subsection{Rule 4: Relevant \textbf{baselines} should always be provided}

It is also a strong recommendation to report relevant performance \textbf{baselines}. A baseline can be selected from different perspectives. To give an example, popular cross-validation curves for the selection of the number of latent variables (LVs) in component models (like PCA or Partial Least Squares (PLS)) are too often shown without a baseline. In this case, a suggested baseline~\cite{camacho2012cross} can be the cross-validated performance criterion (e.g., the sum of squares) of the model with 0 LVs, that is, only considering the preprocessing parameters, like for instance the cross-validated means. Quite often, not a single model variant, regardless the number of LVs, yields an improvement over the 0 LVs baseline. This is a situation in which the selected component model is not an adequate choice, or there is simply no pattern to be found in the data. This situation cannot be detected if a baseline is not provided.

\begin{figure}[htbp]
    \centering
 	\subfigure[]{\includegraphics[width=0.32\textwidth]{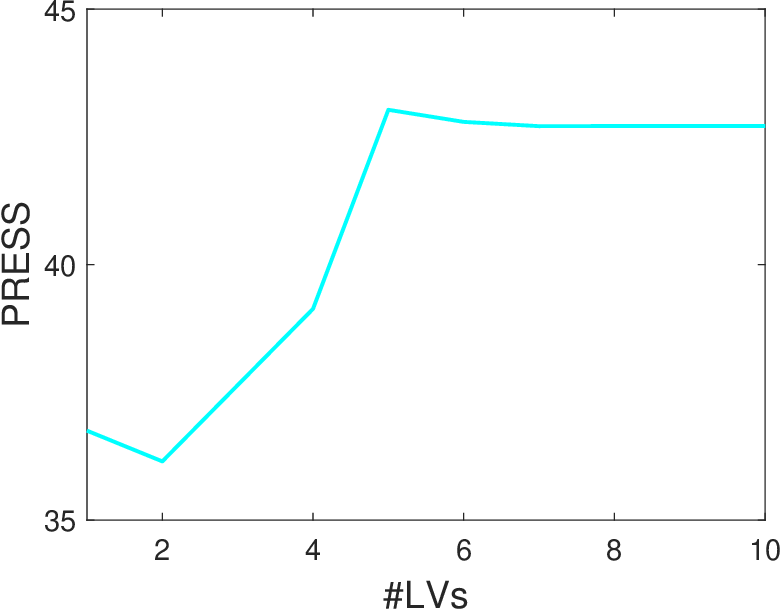}}
 	\subfigure[]{\includegraphics[width=0.32\textwidth]{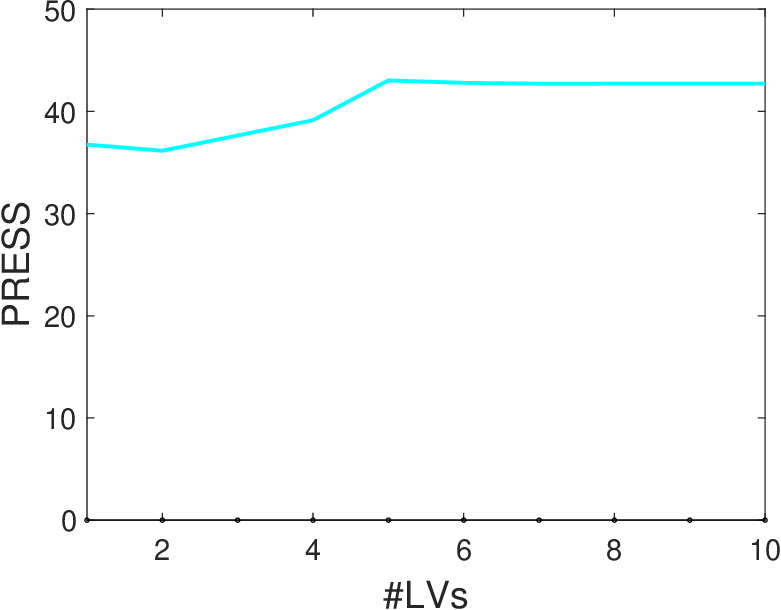}}
 	\subfigure[]{\includegraphics[width=0.32\textwidth]{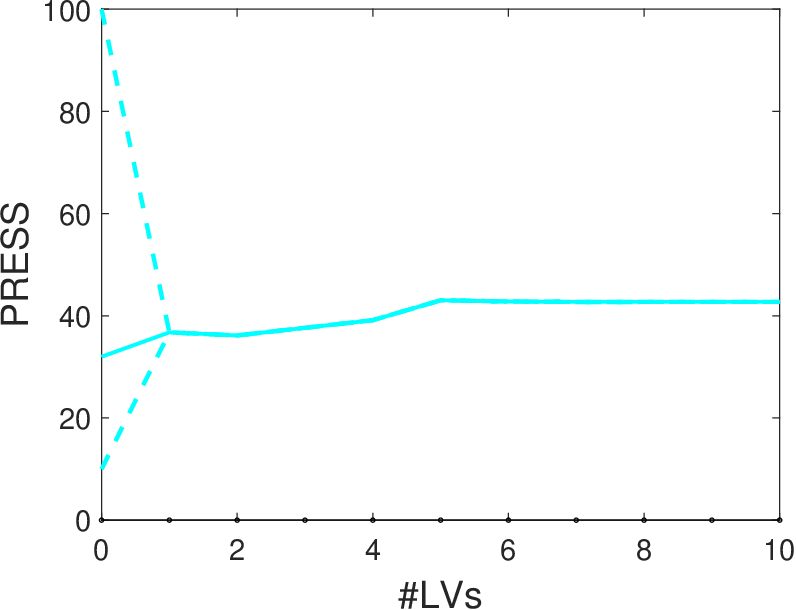}}
    \caption{Cross-validation curve for Partial Least Squares (PLS) in a simulated dataset where $\bX (20 \times 10)$ and $\by (20 \times 1)$ are unrelated.}
    \label{fig:ex2}
\end{figure}

Take the example of Figure \ref{fig:ex2}, with the cross-validation profile of a PLS model where $\bX~(20 \times 10)$ and $\by~(20 \times 1)$  are unrelated. The cross-validation curves are based on the PRESS, defined as:

\begin{equation}
PRESS = \sum_{i=1}^{n} (y_i - \hat{y}_{i})^2
\end{equation}

\noindent where $y_i$ represents each observation in $\by$ and $\hat{y}_{i}$ its corresponding prediction. In cross-validation, $\hat{y}_{i}$ is computed with a model which was not trained with $y_i$. 

Figure \ref{fig:ex2}(a) shows a cross-validation curve with a minimum at two LVs. An obvious baseline we are missing in this first plot is to fix the minimum of the Y scale at PRESS = 0. When we add it in Figure \ref{fig:ex2}(b) we realize that the difference in PRESS between all the models, especially the ones with one LV and two LVs, is very small. Adding the zero LVs baseline, represented by the error of estimating $\by$ with its own (cross-validated) average, also provides useful information. This is illustrated in Figure \ref{fig:ex2}(c), where we also show with dashed lines two hypothetical values for the PRESS at 0 LVs that would dramatically change our interpretation. If the PRESS at zero LVs is 100, the interpretation would be that the model with one or two LVs is actually capturing more than 60\% of the variance in $\by$ in a predictive way. If the PRESS at zero LVs is 10, the interpretation would be that no PLS model has any predictive power. This is actually what the real value of the PRESS at zero LVs (around 30) is implying.

Another form of baseline is represented by \textbf{null examples}. A null example is a 
negative case, fabricated using random number generators while maintaining as much as possible the characteristics of the real data at hand---like the number of rows and columns, group imbalances, selected correlation structures, etc. The null distribution in permutation testing~\cite{anderson2003permutation}, or the null effect size in powercurves~\cite{camacho2023permutation}, for instance, follow this intuition. The resulting performance reflects a baseline, which is often (much) higher than expected~\cite{camacho2022dataset}. Assessing any data pipeline with one or several null examples is an excellent supplement for validation. Null examples are a perfect tool to detect data leakage problems, because they lead to an unexpectedly high performance when such problems exist. 
   
\begin{figure}[htbp]
    \centering
 	\subfigure[$Q^2 = 0.71$]{\includegraphics[width=0.45\textwidth]{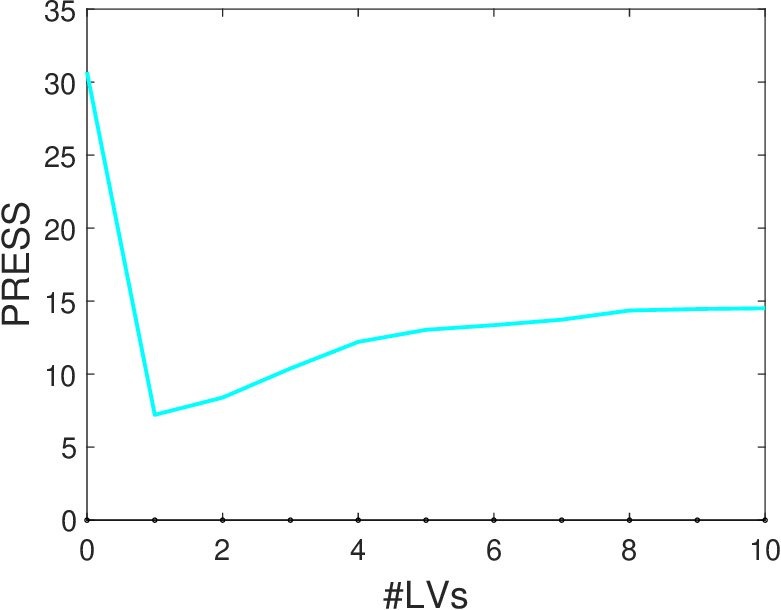}}
 	\subfigure[$Q^2 = -0.15$]{\includegraphics[width=0.45\textwidth]{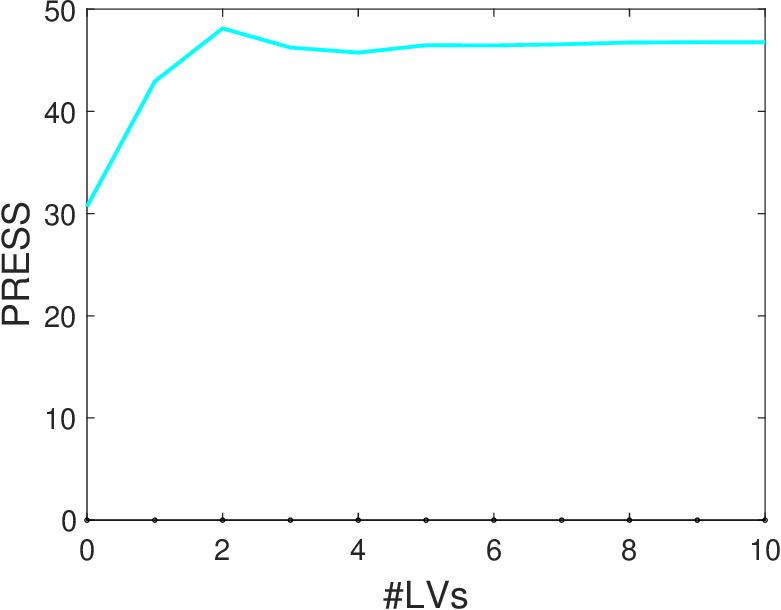}}
    \caption{Cross-validation curve and Goodness of prediction ($Q^2$) of double cross-validation for Partial Least Squares (PLS) in a simulated dataset where $\bX (20 \times 1000)$ and $\by (20 \times 1)$ are unrelated: variable selection performed before validation (a) and variable selection performed within the inner loop (b).}
    \label{fig:ex3}
\end{figure}

Take now the example of Figure \ref{fig:ex3}, similar as the previous example but for $\bX~(20 \times 1000)$ and $\by~(20 \times 1)$. Since both blocks of data are unrelated, this represent a null example of a case where $\bX$ is highly dimensional. A correct validation scheme should reflect the null predictive power of $\bX$ over $\by$. 
Figure \ref{fig:ex3}(a) shows the (incorrect) case where we do variable selection with the whole dataset and then perform double cross-validation. Figure \ref{fig:ex3}(b) shows the (correct) case where we do variable selection within the inner loop of the double cross-validation. In both cases, we present one example of the cross-validation curve in the inner loop, and the final Goodness of prediction ($Q^2$) of the selected model, defined as:

\begin{equation}
Q^2 = 1 - \frac{\sum_{i=1}^{n} (y_i - \hat{y}_{i})^2}{SS_{y}}
\end{equation}

with $SS_{y}$ the sum-of-squares of $\by$. In double cross-validation, $\hat{y}_{i}$ is computed with a model which was neither trained nor selected with $y_i$. 

The null example shows the overly optimistic nature of the incorrect validation scheme in Figure \ref{fig:ex3}(a), where the optimal model has one LV and a fairly good (perceived) prediction performance. The correct validation scheme in Figure \ref{fig:ex3}(b) shows that $\bX$ does not contain information on $\by$. The nice property of a null example is that it is extremely easy to generate, either from random data or by permutation of a real dataset, but very powerful to detect flawed validation schemes. In very complex pipelines, with several steps for preprocessing, feature engineering and modeling, running some permuted examples is an effortless way to double-check that the pipeline is not suffering from any data leakage problem.    
 
\subsection{Rule 5: Comparisons among models should consider both \textbf{statistical and practical significance}}

Favoring the model or data pipeline with the optimal average performance is tricky. If the criterion for performance evaluation has enough resolution---i.e., enough decimal places---there is always a \textbf{'perceived' winner}. However, there may be cases in which models or pipelines are simply \textbf{even in performance}, given the level of uncertainty in the data. For instance, this may be the case when the validation is repeated with different tests sets and the best model changes. For this reason, rather than reporting only average performance metrics, like total sum-of-squares of the error, reporting uncertainty statistics and/or statistical significance is more informative. Taking statistical significance into account at model selection, but in particular for performance evaluation, has practical benefits, because we can always choose the most convenient solution among the similarly performing models. For performance measures that are computed per observation (like the PRESS), we can directly use the distribution of these values across the observations to compute uncertainty or statistical significance\footnote{Care should be taken to properly manage the correlation between performance estimates within and between models}. An alternative way to provide uncertainty metrics is through resampling techniques like bootstrapping~\cite{efron1992bootstrap,mooney1993bootstrapping}. Uncertainty is a critical component of model performance and is necessary for conducting rigorous comparisons across models. Consequently, model uncertainty should be reported to ensure transparency and facilitate future benchmarking.

\begin{figure}[htbp]
    \centering
 	\subfigure[$Q^2 = 0.69~(\pm~0.11)$]{\includegraphics[width=0.31\textwidth]{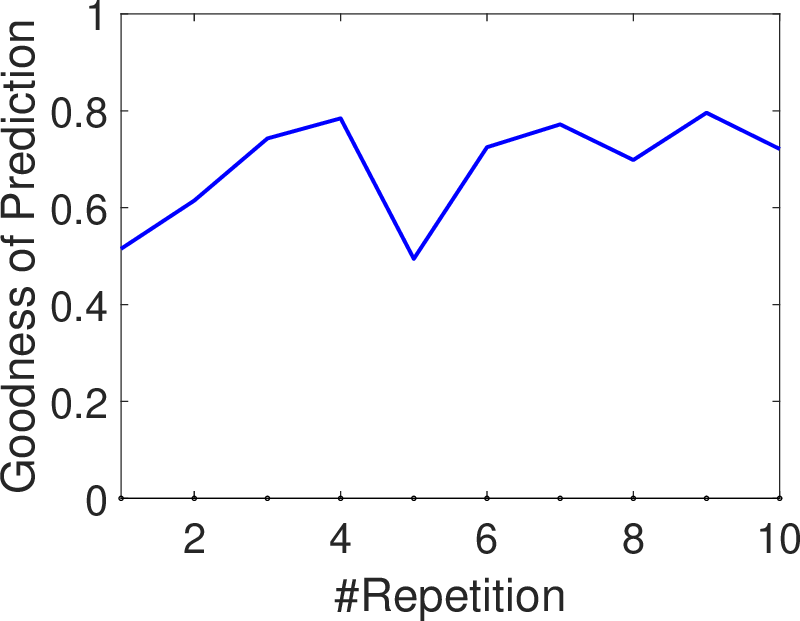}}
 	\subfigure[$Q^2 = 0.75~(\pm~0.04)$]{\includegraphics[width=0.31\textwidth]{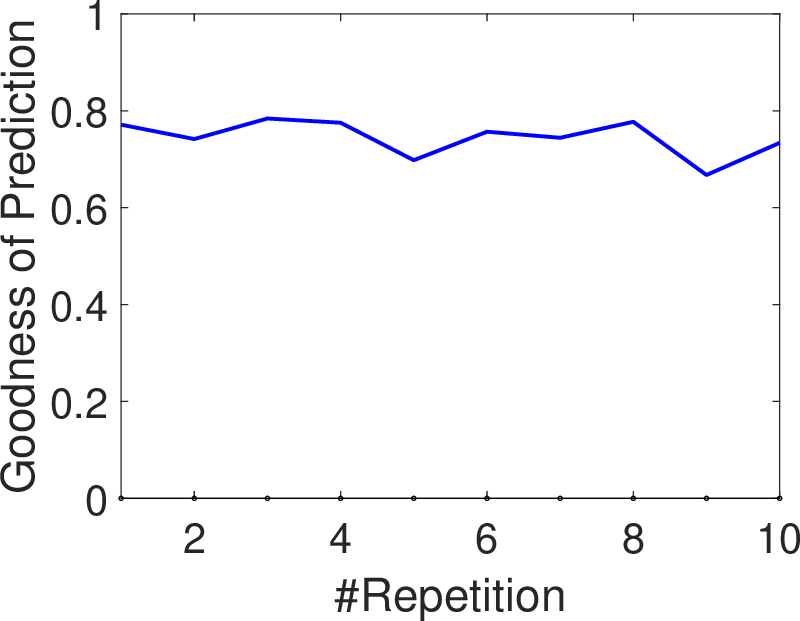}}
 	\subfigure[$Q^2 = 0.77~(\pm~0.05)$]{\includegraphics[width=0.31\textwidth]{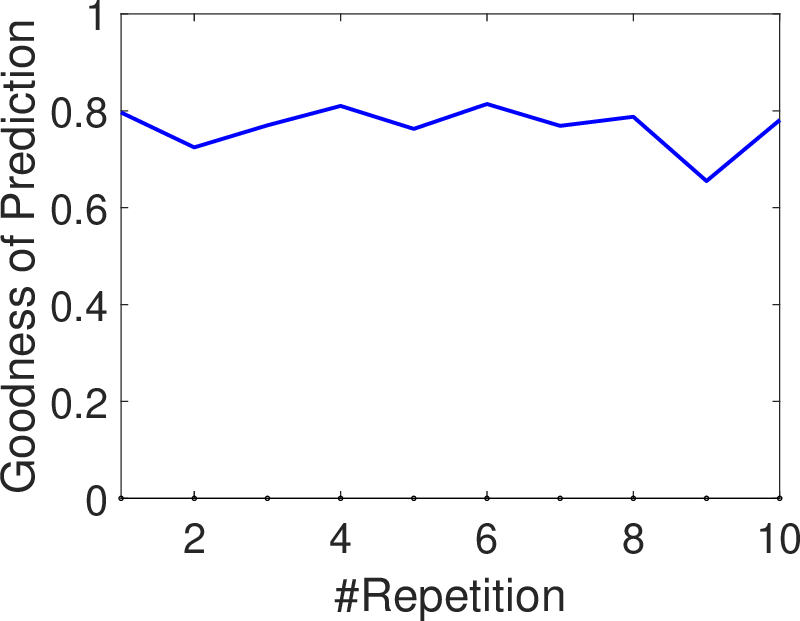}}
 	\subfigure[$Q^2 = 0.67~(\pm~0.12)$]{\includegraphics[width=0.31\textwidth]{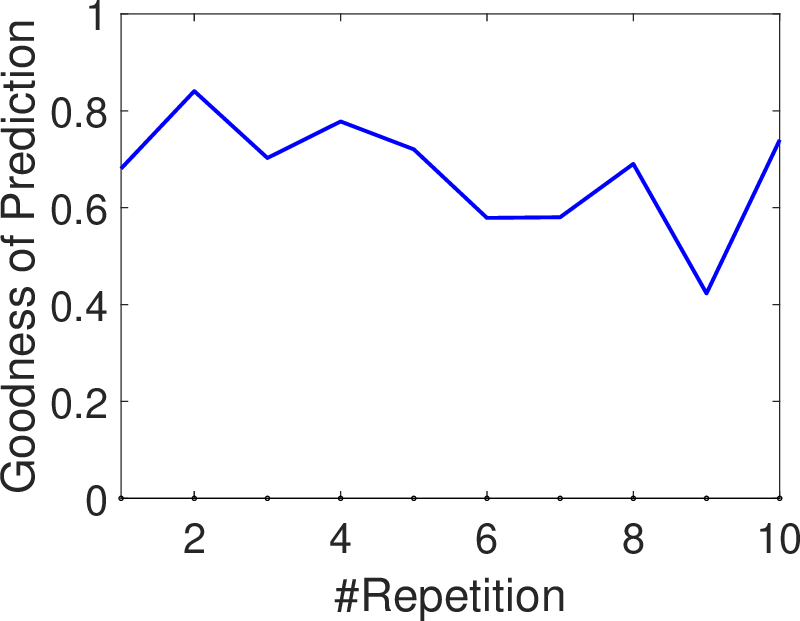}}
 	\subfigure[]{\includegraphics[width=0.31\textwidth]{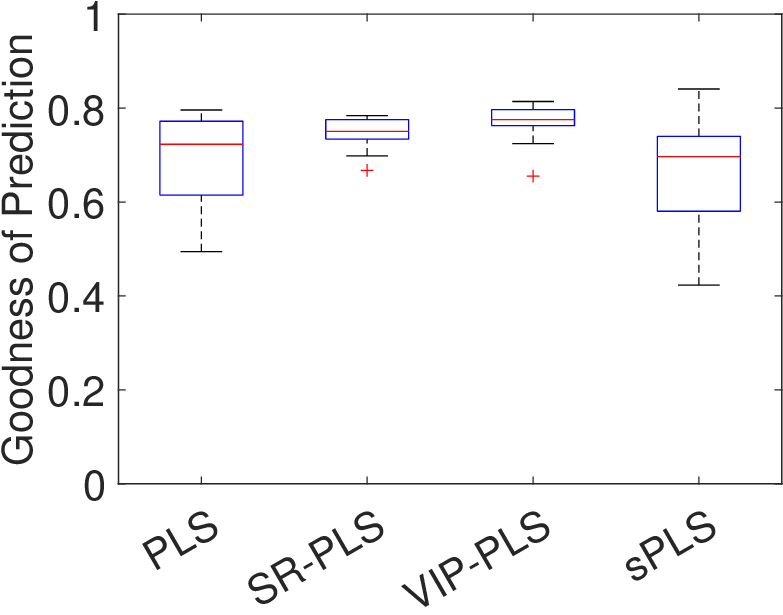}}
    \caption{Repeated $Q^2$ measurements from double cross-validation for simulated data where $\bX~(20 \times 100)$ and $\by~(20 \times 2)$ have a noisy connection between some variables of $\bX$ and $\by$: PLS model (a), PLS model with variable selection with the Selectivity Ratio (b), PLS model with variable selection with the VIP (c), sparse PLS model (d), and comparison of the results with a boxplot (e).}
    \label{fig:ex4}
\end{figure}

Take now the example of Figure \ref{fig:ex4}, based on  simulated data where $\bX~(20 \times 100)$ and $\by~(20 \times 2)$ have a noisy connection between $\by$ and a subset of variables in $\bX$. We perform double cross-validation with ten repetitions for different PLS-based models, with and without variable selection, including the selectivity Ratio (SR)~\cite{rajalahti2009biomarker}, Variable Importance in Projection (VIP)~\cite{wold1993pls} and sparse PLS~\cite{le2015sparse}. Setting aside variability for a moment, let us focus on a single run: repetition one. If we only look at that repetition, our conclusion would be that the performance (from best to worst) is VIP-PLS, SR-PLS, sPLS, and PLS. If we rather take repetition two, we get sPLS, SR-PLS, VIP-PLS, and PLS. And if we take repetition nine: PLS, SR-PLS, VIP-PLS, sPLS. The boxplot provides a more informative comparison, by leveraging the variability in the results: PLS and sPLS provide moderately unstable performance results and SR-PLS and VIP-PLS provide both more stable and better performance. VIP-PLS attains the maximum median, but (very importantly) there is no statistically significant difference between SR-PLS and VIP-PLS (p-value = 0.26), so both models can be considered even since the difference in perceived performance can be the result of chance.      

However, even when statistically significant differences exist among alternative models or data pipelines, the differences may still be negligible from the practical standpoint. Furthermore, there may be additional considerations to make when selecting the 'best' solutions: some models may require impractical data volumes/sizes, processing times and/or energy consumption for training, or make use of impractical information complex or expensive to retrieve, like detailed data labeling for supervised models. An increasingly widespread practice in recent data science literature is to propose a supervised deep learning model even in applications where reliable data labeling may be obtained only manually. This manual labeling may be impractical or even unfeasible, especially for models trained from massive data that require regular updating. Furthermore, an important consideration that is often forgotten is the labeling uncertainty. In most real applications, a certain level of labeling error is always expected, and pushing accuracy above this level is of limited practical value---it is, as a matter of fact, overfitting, but mostly a waste of time and resources.

In the previous example, computational times for the double cross-validation was: 1.3s for PLS, 21.8s for SR-PLS, 16.9s for VIP-PLS and 26.3s for SPLS. Depending on the situation---rates at which models need to be created and validated, expenditures in money and energy associated, etc.---the PLS option may be a more practical solution.

\section{Assessing an Example of Validation Strategy: the Double-check for PLS-DA} \label{Discussion}

Szyma{\'n}ska et al.~\cite{szymanska2012double} propose a very complete validation scheme for metabolomic models with Partial Least Squares Discriminant Analysis (PLS-DA). The proposal combines double cross-validation with permutation testing. The double cross-validation splits the data intro three datasets: model building and test data are separated in the outer loop to guarantee independence between them, and model buiding data is further split into training and validation data in the inner loop, to fit several PLS-DA models with different numbers of LVs. The inner loop then selects an optimal model variant and the generalization performance is estimated with the independent test. The exact same process is repeated many times after permuting the rows in \textbf{Y} or \textbf{X}, leaving the other block untouched, so that a null distribution comprising the spurious relationships in the data is estimated. This is repeated multiple times to assess the uncertainty in the generalization performance estimates. We extended this algorithm to sparse PLS-DA \cite{camacho2015multivariate,jimenez2021pls}, with the addition that both the number of LVs and the sparsity level are to be chosen in the inner loop, and that uncertainty can be used for model selection. 

Let us analyze this validation scheme in the light of the proposed validation rules.

\begin{itemize}

\item{Rule 1:} The test data is made independent by making all model building operations (including preprocessing, selecting the number of LVs and the sparsity level) within the inner loop. 

\item{Rule 2:} The independence level is the one for data splitting, so any technical effect affecting the whole data generation is not considered. If the model is meant to generalize to other batches, technicians or laboratories, or to other analytical instruments, a more developed validation scheme is necessary. Therefore, the approach provides an exploratory solution to find good candidates on metabolites (or alternative variables), but further validation is required for a more extended generalization ability. We should always recognize this limitation when reporting validation results. Furthermore, this validation approach, at least when using random splitting in training, validation and test data, may not be suitable for time series or any other form of data with relationships among observations. This is the case, for instance, of more elaborated studies with repeated measures~\cite{madssen2021repeated}.

\item{Rule 3:} The authors suggest the NMC and the AUROC as more representative statistics for performance evaluation in classification problems with PLS-DA. Yet, one may realize that both metrics treat false positives and negatives equally, and that both may not be fully reliable for severely unbalanced data, so they should be used with caution in the face of these complications.  

\item{Rule 4:} A baseline is provided based on permutation testing, to evaluate if the generalization performance of the model is due to chance or rather statistically significant. This is a very interesting complement to the double cross-validation loop because it allows us to determine whether a given performance result is actually significant for a dataset with specific characteristics: number of rows, number of columns, correlation within \textbf{X} or \textbf{Y} and, very importantly, imbalanced nature.  

\item{Rule 5:} While not implemented in the original publication of Szyma{\'n}ska et al.~\cite{szymanska2012double}, the resampling approach within the cross-validation loop can be leveraged to estimate uncertainty measures in the jackknifing/bootstrap style ~\cite{efron1992bootstrap,efron1981jackknife}, which can be taken into account at model selection. Furthermore, the uncertainty reflected by the repetitions of the double cross-validation can be taken into account when comparing the generalization performance of (s)PLS-DA with other model variants (in the spirit of Figure \ref{fig:ex4}). It should be considered, however, that these variance estimates are not independent and that practical significance should always be taken into account~\cite{mendez2019comparative}.

\end{itemize}

\section{Conclusion} \label{Conclusions}

In this work, I propose a number of rules for model validation that allow us to develop
solid validation schemes in complex scenarios where the risk of overestimating the generalization performance is high. I argue that no validation scheme is completely flawless, but that practical solutions that depend on the model real-life application can be defined. For this reason, it is important to complement the performance generalization results with (1) a description of the limitations of the validation scheme, including the level of independence and completeness of the test set and the suitability and limitations of the performance criteria for a practical case study; and (2) reference information, including baselines, uncertainty measures and statistical and practical significance of the results. I strongly recommend to run the complete validation pipeline with one or several null examples, to assess the Type I error or false positive risk.

\section*{Software}

The examples of the paper can be reproduced using the code at repository \url{https://github.com/josecamachop/RulesOfValidation}. This repository requires the MEDA toolbox v1.10 at \url{https://github.com/codaslab/MEDA-Toolbox}.

\section{Acknowledgements}

I kindly acknowledge the comments by Michael Sorochan Armstrong, Jesús García Sánchez and Daniel Vallejo España, and of the anonymous reviewers. All of them were really helpful to improve the message of the paper. This work was supported by grant no. PID2023-1523010B-IOO (MuSTARD), funded by the Agencia Estatal de Investigación in Spain, call no. MICIU/AEI/10.13039/501100011033, and by the European Regional Development Fund.

\newpage
\afterpage{\clearpage}
\bibliography{Bibliography}
\bibliographystyle{ieeetr}

\end{document}